\documentstyle[12pt,a4,epsf]{article}
\begin{document}
\input psfig
\bibliographystyle{unsrt}
\baselineskip= 18pt
\pagenumbering{arabic}
\pagestyle{plain}
\def\bit{\begin{itemize}}
\def\eit{\end{itemize}}
\def\half{{1\over 2}}
\def\OO{\Omega}
 \def\aa{\alpha}
 \def\bk{{\bf k}}
 \def\bkp{{\bf k'}}
 \def\bqp{{\bf q'}}
 \def\bq {{\bf q}}
 \def\EE{\Bbb E}
 \def\EEx{\Bbb E^x}
 \def\EEo{\Bbb E^0}
 \def\LL{\Lambda}
 \def\PP{\Bbb P^o}
 \def\rr{\rho}
 \def\SS{\Sigma}
 \def\ss{\sigma}
 \def\ll{\lambda}
 \def\dd{\delta}
 \def\ww{\omega}
 \def\ll{\lambda}
 \def\DD{\Delta}
 \def\DDt{\tilde {\Delta}}
 \def\kr{\kappa\lb \LL\rb}
 \def\PPx{\Bbb P^{x}}
 \def\gg{\gamma}
 \def\kk{\kappa}
 \def\tt{\theta}
 \def\lb{\left(}

 \def\rb{\right)}
 \def\prt{\tilde p}
\def\pt{\tilde {\phi}}
 \def\bb{\beta}
 \def\hal{{1\over 2}\nabla ^2}
 \def\bg{{\bf g}}
 \def\bx{{\bf x}}
 \def\bu{{\bf u}}
 \def\bv{{\bf v}}
 \def\by{{\bf y}}
 \def\hag{{1\over 2}\nabla}
 \def\beq{\begin{equation}}
 \def\eeq{\end{equation}}
 \def\cosech{\hbox{cosech}}
 \def\fr{\frac}
 \def\Tr{{\mbox{Tr}}}

\title{Strong Coupling Model for String Breaking on the Lattice.}
\author{I T Drummond  \\
        Department of Applied Mathematics and Theoretical Physics \\
        University of Cambridge \\
        Silver St \\
        Cambridge, England CB3 9EW}
\maketitle
\begin{abstract}
A model for $SU(n)$ string breaking on the lattice is formulated using 
strong coupling ideas. Although necessarily rather crude, the model 
gives an explicit picture of string breaking in the presence of dynamical 
quarks as a mixing phenomenon that involves the string state and a two-meson state. 
A careful analysis, within the model, of the Wilson loop shows that
the evolution of the mixing angle as a function of separation may
obscure the expected crossover effect. If a sufficiently extensive mixing
region exists then an appropriate combination of transition amplitudes can help 
in revealing the effect.

The sensitivity of the mixing region to the values of the meson energy and the 
dynamical quark mass is explored and an assessment made of the dectectibility 
of string breaking in a practical lattice simulation.

\end{abstract}
\vfill
DAMTP-98-35
\pagebreak

\section{Introduction}
Currently much effort in lattice gauge theory simulations
is being devoted to dynamical quarks \cite{DQ1,DQ2,DQ3,DQ4,DQ5,DQ6}. 
For a review see S G\"usken \cite{DQ7}. For computational reasons 
the quarks incorporated in the model are still relatively massive.
This makes it difficult to detect dramatic differences in the 
computed hadron spectrum relative to results obtained in the 
quenched approximation. In order to highlight the effect of dynamical quarks
it is important to investigate phenomena that cannot occur in their absence.
Gauge string breaking is an example of this type of event.
Here we show explicitly that indeed string breaking does occur on the lattice 
and shows up, as expected, as a mixing phenomenon. 

The physical situation envisaged in lattice calculations comprises
a static quark and anti-quark separated by a spatial distance of $R$
lattice units (we mostly set the lattice spacing to unity). The two 
static particles may either support a gauge string that runs between
them or separately bind a quark and an anti-quark to create a two-meson state.
When the string is stretched sufficiently, its energy coincides with 
the energy of the two meson state. In the neighbourhood of this
critical separation the two different physical states mix permitting 
the occurence of transitions between them. 

There are two related issues of importance, namely the size of the range 
in $R$ over which there is substantial mixing and the magnitude of the
energy split induced by mixing. As fractional effects we find that they are
proportional to one another and therefore of essentially the same size.
We show in terms of our model that, for the lowest quark masses and for 
meson excitations that are not too great, the mixing effects of string
breaking may well be detectable. If the meson masses are too great and/or  the 
quarks too heavy then the phenomemnon will be undetectable on the lattice.

Because it is based on strong coupling ideas the results of our model
are provisional and require interpretation if they are to be applied 
to actual simulations. Nevertheless the model is very suggestive and 
identifies crucial aspects of the mixing phenomenon to which the results 
are sensitive.

\section{Strong Coupling Model}

We formulate the model by first recalling the well known rules for evaluating 
simple graphs in the strong coupling limit of $SU(n)$ gauge theory \cite{CREUTZ, MonMun}. 
They are as follows:
\bit
\item[1.] A factor of 
$$
\left(\fr{\bb}{2n^2}\right)
$$
for each plaquette.  
\item[2.] A factor of
$$
2\kk\left(\fr{1+\gg.e}{2}\right) 
$$
for each Wilson quark line in the direction of the unit vector $e$~.
Here $\kk$ is the standard quark hopping parameter.
\item[3.] A factor of $(-1)$ for each internal quark loop.
\item[4.] A factor of $1/n$ for each internal quark loop.
\item[5.] A trace over the spin matrix factors for each internal quark loop.
\eit
\subsection{Simple String Model}

In the absence of dynamical quarks our model for the correlation function 
of the string of length $R$ over an imaginary time interval $T$ is the standard 
$R\times T$ Wilson loop. In leading strong coupling approximation the above 
rules give for this string-string
propagator
\beq
{\cal G}_{SS}(T)=\left[\left(\fr{\bb}{2n^2}\right)^R~\right]^T~~.
\eeq
This tells us immediately that the energy of the string state is $V(R)$
where
\beq
V(R)=\ss R~~,
\eeq
and $\ss$ is the dimensionless string tension given by
\beq
\ss=-\log\left(\fr{\bb}{2n^2}\right)~~.
\eeq

Note that we have omitted the spin degrees of freedom of the static quarks.
They play no r\^ole in the model.

\subsection{Model for Mesons}

The simple model for mesons suggested by the strong coupling limit,
is one in which a light quark propagates along a static (anti-)quark
line. See Fig 1~. Using the rules, we find that this meson propagator 
has the structure
\beq
g(T)=\fr{1+\gg_0}{2}\left(2\kk\right)^T~~.
\eeq
Of course we retain the spin degrees of freedom of the light quarks.
They do play a r\^ole in the model. 

The propagator for two mesons moving independently each bound to
its static quark is
\beq
g^{(1)}(T)\otimes g^{(2)}(T)=
        \left(\fr{1+\gg_0}{2}\right)^{(1)}\otimes
        \left(\fr{1-\gg_0}{2}\right)^{(2)}\left((2\kk)^2\right)^T~~.
\eeq

In fact, as will become clear below, only a particular combination
of quark-anti-quark spin wave functions enters the calculation.
It is obtained by completing the light quark loop with the matrices
\beq
\fr{1\pm\gg_1}{2}~~,
\eeq
and including a factor of $(-1)$~. This two-meson propagator is
\beq
{\cal G}_{MM}(T)=
-\Tr\left(\fr{1-\gg_1}{2}\right)g^{(1)}(T)\left(\fr{1+\gg_1}{2}\right)g^{(2)}(T)
   =\fr{1}{2}\left((2\kk)^2\right)^T~~.
\label{MESONS}
\eeq

Finally we will give the light quarks,
when bound in a meson, a hopping parameter $\kk'$ that may be 
distinct from the value $\kk$ that holds elsewhere in the diagrams of the model.
Eq(\ref{MESONS}) should be modified therefore by the replacement $\kk\rightarrow\kk'$~.
The essential point is that we are treating the energy, $E_M$, of the mesons as a parameter.
This is reasonable since the energy of the mesons is an important number in the
string breaking process and a simple version of the hopping parameter expansion
cannot reflect the full dynamics of the meson bound state. 
We asume then
\beq
(2\kk')^2=e^{-E_M}~~,
\eeq
where $E_M$ is the combined energy of the two bound mesons.

\section{String Breaking}

Our model for string breaking involves summing over planar graphs
that incorporate transitions between the string and meson states
described above. We arrive at the rules for computing these graphs 
by considering the case of string-to-string propagation represented by 
Fig 2. From the rules we see that the corresponding contribution to the Green's
function is
$$
\left[\left(\fr{\bb}{2n^2}\right)^R\right]^{T_1}\left[\left(2\kk'\right)^2\right]^{T_2}
   \left[\left(\fr{\bb}{2n^2}\right)^R\right]^{T_3}\left[\left(2\kk'\right)^2\right]^{T_4}
     \left[\left(\fr{\bb}{2n^2}\right)^R\right]^{T_1}
$$
\beq
 \left(-\fr{t}{n}\right)\left[\left(2\kk\right)^2\right]^R
 \left(-\fr{t}{n}\right)\left[\left(2\kk\right)^2\right]^R~~,
\eeq
where 
\beq
t=\Tr\left(\fr{1+\gg_0}{2}\right)\left(\fr{1+\gg_1}{2}\right)
     \left(\fr{1-\gg_0}{2}\right)\left(\fr{1-\gg_1}{2}\right)=-\fr{1}{2}~~.
\eeq
The factors involving $\kk$ (as opposed to $\kk'$) are associated with the 
{\it vertical} sides of the quark loops. Note that the trace of the internal quark 
loop is equivalent to the quark-anti-quark spin projection mentioned previously.

We reinterpret the structure of such diagrams as follows:

\bit
\item[1.] The factor 
$$
\left(\fr{\bb}{2n^2}\right)^R
$$
propagates the string by one time step.
\item[2.] The factor
$$
\left(2\kk'\right)^2
$$
propagates the two-meson state by one time step.
\item[3.] The factor
$$
\fr{1}{\sqrt{2n}}\left(2\kk\right)^R
$$
is associated with the transition from string to two-meson state
and vice-versa.
\eit

At any stage we may view the system as being in either a string
or a two-meson state. To describe the transition from an initial time
to time $T$ we need a $2\times 2$ matrix of transition amplitudes
\beq
G(T)=\left(\begin{array}{cc}G_{SS}(T)&G_{SM}(T)\\ G_{MS}(T)&G_{MM}(T)\end{array}\right)~.
\eeq
If for convenience, we introduce the parameters
\beq
a=\left(\fr{\bb}{2n^2}\right)^R~~,~~~~~~~~b=\left(2\kk'\right)^2~~,
   ~~~~~~~~c=\fr{1}{\sqrt{2n}}\left(2\kk\right)^R
\eeq
then the above stepping procedure can be represented by
\beq
G(T+1)=A\left(\begin{array}{cc}G_{SS}(T)&G_{SM}(T)\\ G_{MS}(T)&G_{MM}(T)\end{array}\right)~~,
\eeq
where the matrix $A$ is given by
\beq
A=\left(\begin{array}{cc}a&ac\\ bc&b\end{array}\right)~~.
\eeq
If for definiteness  we set 
\beq
G(0)=1~~,
\eeq
then
\beq
G(T)=\left(A\right)^T~~.
\eeq
The matrix $A$ is not symmetrical but can be expressed in terms of a symmetric
matrix $B$ in the form
\beq
A=DBD^{-1}~~,
\eeq
where $D$ is the diagonal martix
\beq
D=\left(\begin{array}{cc}\sqrt{a}&0\\0&\sqrt{b}\end{array}\right)~~,
\eeq
and
\beq
B=\left(\begin{array}{cc}a &\sqrt{ab}~c\\ \sqrt{ab}~c &b\end{array}\right)~~.
\eeq
The eigenvalues of $B$ are
\beq
\ll_{\pm}=\fr{1}{2}\left\{(a+b)\pm\sqrt{(a-b)^2+4abc^2}\right\}~~.
\eeq
The corresponding eigenvectors are
\beq
\chi_{+}=\left(\begin{array}{c}\cos\theta \\ \sin \theta \end{array}\right)~~,~~~~~~~~
\chi_{-}=\left(\begin{array}{c}-\sin\theta \\ \cos \theta \end{array}\right)~~,
\eeq
where
\beq
\tan\theta=\fr{-(a-b)+\sqrt{(a-b)^2+4abc^2}}{2\sqrt{ab}~c}~~.
\eeq
If we construct the othogonal matrix
\beq
O=\left(\begin{array}{cc}\cos\theta&-\sin\theta\\\sin\theta&\cos\theta\end{array}\right)~~,
\eeq
then
\beq
B=O\Lambda O^{-1}
\eeq
where
\beq
\Lambda=\left(\begin{array}{cc}\ll_{+}&0\\0&\ll_{-}\end{array}\right)~~.
\eeq

It follows immediately that for any $T$
\beq
A^T=DO\left(\begin{array}{cc}\ll_{+}^T&0\\0&\ll_{-}^T\end{array}\right)O^{-1}D^{-1}~~.
\eeq
This shows clearly that the propagating eigenstates are mixtures of the string
and two-meson states determined by the mixing angle $\theta$~. The corresponding 
eigen-energies are
\beq
E_{\pm}=-\log\ll_{\pm}~~.
\eeq
Note that $E_{+}$ is the {\it lower} of the two energies.

\section{Wilson Loop}

We can now use the above results to elucidate the behaviour of the 
Wilson loop provided we pay attention to end conditions. The Wilson loop 
begins with a factor $a$~. The sum of all such diagrams of length $T$ are
obtained as the components of
\beq
A^{T-1}\left(\begin{array}{c}a\\0\end{array}\right)~~.
\label{STRING}
\eeq
The Wilson loop itself is obtained as
\beq
W(R,T)={\cal G}_{SS}(T)=(1,0)DO\Lambda^{T-1}O^{-1}D^{-1}\left(\begin{array}{c}a\\0\end{array}\right)~~.
\eeq
That is
\beq
W(R,T)=a\left(\cos^2\theta~\ll_{+}^{T-1}+\sin^2\theta~\ll_{-}^{T-1}\right)~~.
\eeq
The Wilson loop, therefore, sees both exponentials 
\beq
e^{-E_{\pm}T}=\ll_{\pm}^T~~.
\eeq
In principle, the the lower energy exponential should dominate the asymptotic behaviour.
In practice, what is observed will be influenced by the behaviour of the coefficients of the 
exponential contributions through their dependence on the mixing angle.

Noting that
\beq
\sin^2\theta=\fr{-(a-b)+\sqrt{(a-b)^2+4abc^2}}{2\sqrt{(a-b)^2+4abc^2}}~~,
\eeq
we see that when $V(R)<<E_M$ then $a>>b$ and we have
\beq
\theta\simeq 0~~,~~~~~~~~\ll_{+}\simeq a~~,~~~~~~~~\ll_{-}\simeq b~~,
\eeq
with the result that $E_{+}\simeq V(R)$ and $E_{-}\simeq E_M$~. Under these circumstances 
the coupling to the state with energy $E_{-}$ will vanish and only the exponential 
associated with $E_{+}$ will be observed as expected. When $R$ is chosen so that
$V(R)>>E_M$ then $b>>a$ and we have
\beq
\theta\simeq \fr{\pi}{2}~~,~~~~~~~~\ll_{+}\simeq b~~,~~~~~~~~\ll_{-}\simeq a~~,
\eeq
with the result that $E_{+}\simeq E_M$ and $E_{-}\simeq V(R)$~. Under these circumstances the
coupling to the state with energy $E_{+}$ will vanish and only the exponential 
associated with $E_{-}$ will be observed. That is, both above and below the crossover
point, only the 
{\it original} string behaviour 
$$
e^{-V(R)T}~~,
$$
will be observed. The movement of the mixing angle therefore obscures the crossover phenomenon 
in which the string energy is expected to be bounded by the energy of the
two-meson state. As is clear from the above analysis this does not happen.

To see mixing effects directly in the Wilson loop we must fix $R$ near the 
critical crossover value $R_c$, for which $V(R)=E_M$ and hence $a=b$~. 
That is
\beq
R_c=\fr{E_M}{\ss}~~.
\eeq
At this point $\theta=\pi/4$ and 
\beq
W(R,T)=\fr{a}{2}\left(\ll_{+}^{T-1}+\ll_{-}^{T-1}\right)~~,
\eeq
so the two exponentials appear with equal strength. In order to separate
them in a measurement it must be possible to measure their difference.
We find 
\beq
\ll_{\pm}=b(1\pm c)~~.
\eeq
So approximately for small $c$
\beq
E_{\pm}=-\log\ll_{\pm}\simeq E_M(1\mp c)~~,
\eeq
that is
\beq
\fr{\Delta E}{E_M}=\fr{2c}{\ss R_c}~~.
\eeq

We can define the mixing region as the range of $R$ for which 
both $\sin^2\theta$ and $\cos^2\theta$ are significantly different 
from unity. The range of the mixing region in $R$ can be estimated as 
\beq
\fr{\Delta R}{R_c}=\left(\fr{\pi}{2R}\fr{dR}{d\theta}\right)_{R=R_c}~~.
\eeq
It is closely related to the mixing energy we find
\beq
\fr{\Delta R}{R_c}=\pi\fr{\Delta E}{E_M}~~.
\eeq
We estimate the numerical significance of these results below.

\section{String-Meson Transition}

The string-meson transition amplitude can computed from eq(\ref{STRING}) as
\beq
{\cal G}_{MS}(T)=(0,1)A\left(\begin{array}{c}a\\0\end{array}\right)~~.
\eeq
This gives
\beq
{\cal G}_{MS}(T)=\sqrt{ab}~\sin\theta\cos\theta\left(\ll_{+}^{T-1}-\ll_{-}^{T-1}\right)~~.
\eeq
Because of the presence of the factor $\sin\theta\cos\theta$ this transition
amplitude vanishes outside the mixing region. In this region, however, it is possible
to detect the presence of the dominant energy provided the two energies are sufficiently
well separated. However if the two eigen-energies are only weakly separated then
the amplitude will experience a further suppression because the two contributions have
opposite signs that give rise to a cancellation. This is consistent with the results 
of the previous section on the Wilson loop and fits in with the obvious idea that if the primitive
transition process vanishes then the transition amplitude itself vanishes. 

\section{Meson-Meson Transition}

The meson-meson transition involves a sum over amplitudes that begin and end
with a factor $b$~. They can be evaluated as
\beq
{\cal G}_{MM}(T)=(0,1)A\left(\begin{array}{c}0\\b\end{array}\right)~~.
\eeq
The result is 
\beq
{\cal G}_{MM}(T)=b\left(\sin^2\theta~\ll_{+}^{T-1}+\cos^2\theta~\ll_{-}^{T-1}\right)
\eeq
Clearly this amplitude shows properties similar to that for string-string
transitions in that on both sides of the mixing region in $R$ only the
meson energy is detected in the $T$-dependence of the correlator. However the
result suggests that when $R$ is within the mixing region, the combination 
$b~{\cal G}_{SS}(T)+a~{\cal G}_{MM}(T)$ will show a reduced influence from the
mixing angle and will allow the best chance of tracking both $E_{+}$ and $E_{-}$
individually from the resulting $T$-dependence. Of course this requires
favourable circumstances in which the mixing region is sufficiently large
and the energy separation of the mixed states is sufficiently great.

\section{Interpretation of Results}

In order to use the results of our analysis we will make the assumption
that the parameters of the model can be interpreted directly in terms of the appropriate
physical parameters of an actual simulation rather than as bare lattice parameters. 
We have already anticipated this approach when we introduced $\kk'$ as a way
of parametrising the meson energy $E_M$~. 
In the same spirit we will interpret the other hopping parameter $\kk$ in terms of 
the light quark mass, $m_q$, and require
\beq
2\kk=e^{-m_q}~~.
\eeq
We therefore have the identifications
\beq
a=e^{-\ss R}~~,~~~~~~~~b=e^{-E_M}~~,~~~~~~~~c=\fr{1}{\sqrt{n}}e^{-m_qR}~~.
\eeq
With these identifications we then see that the fractional energy shift due to mixing is
\beq
\fr{\Delta E}{E_M}=\sqrt{\fr{2}{3}}\fr{e^{-m_qR_c}}{E_M}~~,
\eeq
and the mixing range in $R$ is
\beq
\Delta R=\pi\sqrt{\fr{2}{3}}\fr{e^{-m_qR_c}}{E_M}R_c~~.
\eeq
We have now specialised to $SU(3)$~. 

If we recall that $R_c=E_M/\ss$ then we see from this formula, through its 
dependence on $R_c$, that the mixing range is very sensitive to both the meson 
energy, $E_M$ and to the quark mass, $m_q$~. In fact
\beq
\Delta R=\pi\sqrt{\fr{2}{3}}\fr{e^{-m_qE_M/\ss}}{\ss}~~.
\eeq

\begin{table}
\begin{center}
\begin{tabular}{|l|l|l|l}\hline
$m_q$~(GeV)&$\Delta E/E_M$&$\Delta R$\\\hline
~~~.1&.7077&18.3\\
~~~.2&.4089&10.6\\
~~~.3&.2363&6.1\\
~~~.4&.1365&3.5\\
~~~.5&.0789&2.0\\
\hline
\end{tabular}
\end{center}
\caption{These results are for $a^{-1}=1.5$ GeV, $\ss^{1/2}=0.427$ GeV,
$E_M=1.0$ GeV. The critical distance is $R_c=8.22$ lattice spacings.}
\label{tab:results1}
\end{table}

\begin{table}
\begin{center}
\begin{tabular}{|l|l|l|l}\hline
$m_q$~(GeV)&$\Delta E/E_M$&$\Delta R$\\\hline
~~~.1&.2044&10.6\\
~~~.2&.0682&3.5\\
~~~.3&.02295&1.2\\
~~~.4&.0076&0.4\\
~~~.5&.0025&0.1\\
\hline
\end{tabular}
\end{center}
\caption{These results are for $a^{-1}=1.5$ GeV, $\ss^{1/2}=0.427$ GeV,
$E_M=2.0$ GeV. The critical distance is $R_c=16.45$ lattice spacings.}
\label{tab:results2}
\end{table}

In order to obtain numerical estimates we consider a range of reasonable 
values appropriate to practical simulations for these parameters. The 
results are shown in Tables 1\& 2~.  These results show directly the 
sensitivity of the physical picture to the energy of the two-meson state
and to the quark mass. For an inverse lattice spacing of 1.5 GeV 
and a two-meson energy of 1 GeV, we obtain
a critical separation $R_c$ of roughly 8 lattice spacings. At the rather 
light value of 100 MeV for the quark mass we see that the mixing region 
is large and covers a range of more than twice this critical distance.
For such circumstances we might expect to see the effects of mixing clearly.
As the quark mass increases in size the mixing region shrinks rapidly so that
by about 400 MeV it has reduced to a barely observable range of just
over 1 lattice spacing either side of the critical separation. 
When the two-meson energy is increased to 2 GeV only the lightest quark mass
of 100 MeV yields a mixing region of any size. At the lower two-meson energy
the the results suggest that there would be a reasonably detectable
energy difference for the lower quark masses. At the larger two-meson energy
only the lightest quark mass yields a substantial mass difference
in the mixing region.

\section{Conclusions}

We have formulated a simple model of string breaking in terms of 
strong coupling ideas. The model shows explicitly string breaking 
as a mixing phenomenon between string and two-meson states.
Because of the nature of the dependence of the Wilson loop
on the angle of mixing between these two states, we do not expect to
see the crossover phenomenon for the string tension in this amplitude. Instead
the original string tension should be observed on {\it both} sides of the
mixing region in $R$~. In a similar way the meson-meson correlator
will show only the meson energy on both sides of the the mixing region.
The string-meson transition amplitude vanishes outside the mixing
region and suffers a suppression inside the region because the
eigen-energy exponentials enter with opposite signs. This amplitude
should therefore show at best only a weak signal in a simulation
unless the mixing region is sufficiently large and the eigen-energies
sufficiently distinct. The model suggests that a particular combination 
of the string-string and meson-meson amplitudes would show a reduced dependence 
on the mixing angle and give the best chance of tracking the two eigen-energy
exponentials through the mixing region. In a practical simulation 
this combination would have to be sought empirically.

To observe string breaking it is necessary to be able to measure amplitudes
in the mixing region with sufficient accuracy. For this to be possible
a simple interpretation of the model in terms of realistic simulation
parameters suggests that it will only be possible if the meson energy is
not too high and the quark mass is sufficiently low. More precisely
we would expect that at the critical value of the static quark
separation, $R_c$, we have $V(R_c)\leq 1.0$ GeV and $m_q\leq 200$ MeV.
The sensitivity of the mixing region to the value of the light quark mass, 
predicted by the model, may yield interesting results on string breaking 
if exploited systematically in practical simulations.

\section*{Acknowledgements}
The author would like to thank Professor Y Iwasaki and Professor A Ukawa
for the hospitality of the Research Center for Computational Physics,
University of Tsukuba, where this work was carried out. He is grateful for
useful and informative discussions with Akira Ukawa, Kazuyuki Kanaya, Ruedi Burkhalter,
Hugh Shanahan and Bengt Petersson.

\newpage
{~~~~~~~~~~~~~~~~~~~~~~~~~~~~~~~}
\vskip 20 truemm
\begin{figure}[htb]
\begin{center}\leavevmode
\epsfxsize=10 truecm\epsfbox{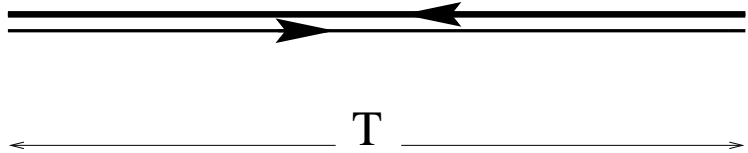}
\end{center}
\caption[]{A dynamical quark (light line) bound to a static quark (heavy line).
There is closure at the ends of the lines and an implied trace for the 
gauge degrees of freedom but not for the light quark spin degrees of freedom.}
\label{figure:F1}
\end{figure}
\vskip 20 truemm
\begin{figure}[htb]
\begin{center}\leavevmode
\epsfxsize=10 truecm\epsfbox{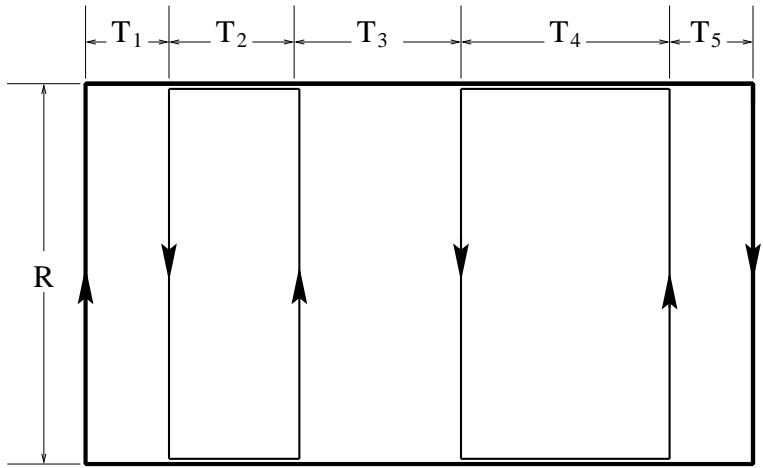}
\end{center}
\caption[]{Wilson loop (heavy line) containing internal quark loops (light lines).}
\label{figure:F2}
\end{figure}


\begin{thebibliography}{99}
\bibitem{DQ1} SESAM and $\mbox{T}_\chi$L Collaborations: G S Bali {\it et al}, 
Nucl. Phys. B (Proc. Suppl.) 63A-C (1998) 209-211
\bibitem{DQ2} D Mawhinney Nucl. Phys. B (Proc. Suppl.) 63A-C (1998) 212-214
\bibitem{DQ3} MILC Collaboration: C Bernard {\it et al}, 
Nucl. Phys. B (Proc. Suppl.) 63A-C (1998) 215-217
\bibitem{DQ4} SESAM and $\mbox{T}_\chi$L Collaborations: H Hoeber {\it et al}, 
Nucl. Phys. B (Proc. Suppl.) 63A-C (1998) 218-220
\bibitem{DQ5} CP-PACS Collaboration: S Aoki {\it et al}, 
Nucl. Phys. B (Proc. Suppl.) 63A-C (1998) 221-226
\bibitem{DQ6} M Talevi, for the UKQCD Collaboration, 
Nucl. Phys. B (Proc. Suppl.) 63A-C (1998) 209-211
\bibitem{DQ7} S G\"usken, Nucl. Phys. B (Proc. Suppl.) 63A-C (1998) 16-21
\bibitem{CREUTZ} M Creutz, ``Quarks and gluons and lattices'', Cambridge
Monographs on Mathematical Physics, Cambridge University Press, 1983
\bibitem{MonMun} I Montvay and G M\"unster, ``Quantum Fields on a Lattice'',
Cambridge Monographs on Mathematical Physics, Cambridge University Press, 1994

\end{thebibliography}
\end{document}